\documentclass[aps,prl,amsmath,amssymb,twocolumn,floatfix]{revtex4-1}
\usepackage{graphicx}
\usepackage{dcolumn}
\usepackage{bm}

\begin{document}

\title{Intrinsic Low-Temperature Magnetism in SmB$_6$}

\author{S.~Gheidi,$^1$ K.~Akintola,$^1$ K. S.~Akella,$^1$ A. M.~C\^{o}t\'{e},$^{1,2}$ S. R.~Dunsiger,$^{1,3}$ C.~Broholm,$^{4,5,7}$ W. T.~Fuhrman,$^6$ S. R.~Saha,$^6$ J.~Paglione,$^{6,7}$, and J. E.~Sonier$^{1,7}$}

\affiliation{$^1$Department of Physics, Simon Fraser University, Burnaby, British Columbia V5A 1S6, Canada \\
$^2$Kwantlen Polytechnic University, Richmond, British Columbia V6X 3X7, Canada \\
$^3$Centre for Molecular and Materials Science, TRIUMF, Vancouver, British Columbia V6T 2A3, Canada \\  
$^4$Institute for Quantum Matter and Department of Physics and Astronomy, The Johns Hopkins University, Baltimore, Maryland, 21218, USA \\
$^5$Department of Materials Science and Engineering, The Johns Hopkins University, Baltimore, Maryland, 21218, USA \\
$^6$Center for Nanophysics and Advanced Materials, Department of Physics, University of Maryland, College Park, Maryland 20742, USA \\
$^7$Canadian Institute for Advanced Research, Toronto, Ontario M5G 1Z8, Canada}

\date{\today}
\begin{abstract}
By means of new muon spin relaxation ($\mu$SR) experiments, we disentangle extrinsic and intrinsic sources of low-temperature bulk magnetism in 
the candidate topological Kondo insulator (TKI) SmB$_6$. Results on Al-flux grown SmB$_6$ single crystals are compared to those 
on a large floating-zone grown $^{154}$Sm$^{11}$B$_6$ single crystal in which a 14~meV bulk spin exciton has been detected by inelastic neutron scattering (INS).
Below $\sim \! 10$~K we detect the gradual development of quasi-static magnetism due to rare-earth impurities and Sm vacancies. 
Our measurements also reveal two additional forms of intrinsic magnetism: 
1) underlying low-energy ($\sim \! 100$~neV) weak magnetic moment ($\sim 10^{-2}$~$\mu_{\rm B}$) fluctuations similar to
those detected in the related candidate TKI YbB$_{12}$ that persist down to millikelvin temperatures, and 2) magnetic fluctuations consistent with a 2.6~meV 
bulk magnetic excitation at zero magnetic field that appears to hinder surface conductivity above $\sim \! 4.5$~K. We discuss potential origins of
the magnetism.               
\end{abstract}

\maketitle
In recent years there has been a concerted effort to determine whether the intermediate-valence compound SmB$_6$ is a strongly correlated three-dimensional (3-D) topological 
insulator (TI). A 3-D TI possesses an insulating bulk and topologically-protected metallic surface states, where the electron spin is locked perpendicular 
to the crystal momentum by strong spin-orbit coupling \cite{Hasan:10,Qi:11}. What makes SmB$_6$ so different from known 3D TIs \cite{Hsieh:2008,Xia:2009,Chen:2009} 
is that it hosts an unconventional insulating bulk gap that forms due to Kondo hybridization of itinerant Sm $5d$ electrons with localized Sm $4f$ states.
As expected for a 3-D TI, experiments on SmB$_6$ have established that metallic surface states dominate the electrical transport 
below $T \! \sim \! 5$~K \cite{Kim:13,Wolgast:13,Syers:15} and a truly insulating bulk exists down to at least 2~K \cite{Eo:19}. Yet there is ongoing debate as 
to whether the surface states are of topological origin. While scanning tunneling microscopy experiments support the existence of heavy in-gap topological 
Dirac fermion states at the predicted locations in the surface Brillouin zone \cite{Pirie:18}, Dirac points have yet to be clearly observed by angle resolved 
photoemission spectroscopy (ARPES) \cite{Xu:14a}. Furthermore, spin-polarized ARPES experiments aimed at determining whether the surface states have the topological 
property of spin-momentum locking have reached very different conclusions \cite{Xu:14b,Hlawenka:18}. 
 
While a true Kondo insulator is non-magnetic, low-temperature magnetism is clearly present in SmB$_6$. There is a field-dependent divergence of the temperature 
dependence of the bulk magnetic susceptibility $\chi(T)$ below $\sim \! 15$~K, originally attributed to bare Sm$^{3+}$ ($4f^5$) magnetic moments, but 
later ascribed to paramagnetic rare-earth impurities incorporated during sample growth \cite{Roman:97,Gabani:02}. Magnetic impurities, which can
destroy the topological protection of surface states by breaking time-reversal symmetry, are also responsible for a large field-induced enhancement of the thermal 
conductivity \cite{Boulanger:18}.

There is also evidence for intrinsic bulk magnetic excitations in SmB$_6$. A 14~meV bulk spin exciton has been detected by INS 
\cite{Alekseev:93,Alekseev:95,Fuhrman:15} and there are reports of lower-energy bulk magnetic excitations potentially relevant to the temperature range over 
which SmB$_6$ exhibits topological behavior. Nuclear magnetic resonance (NMR) measurements indicate the 
existence of intrinsic bulk magnetic in-gap states separated from the conduction band by a 2.6~meV gap that shrinks with increasing field \cite{Caldwell:07} 
and $\mu$SR experiments detect slowly fluctuating internal magnetic fields that persist down to at least 0.02~K \cite{Biswas:14,Biswas:17,Akintola:17}. 
Recently, muon Knight shift measurements on SmB$_6$ at $H\! = \! 60$~kOe have provided evidence for bulk magnetic excitations governed by an $\sim \! 1$~meV 
thermal activation energy \cite{Akintola:18}. While an additional $\lesssim \! 1$~meV spin exciton is predicted \cite{Knolle:17}, the magnetic
excitations at $H\! = \! 60$~kOe may derive from the zero-field-extrapolated 2.6~meV magnetic in-gap states detected by NMR. A $\lesssim \! 2.6$~meV 
spin exciton may hinder topological behavior via spin-flip scattering of the metallic surface states \cite{Kapilevich:15,Arab:16}. 
Surprisingly, however, no collective magnetic excitation has been detected by INS below 14~meV \cite{Fuhrman:17}. 

Here we report new zero-field (ZF) and longitudinal-field (LF) $\mu$SR measurements of SmB$_6$ that enable us to disentangle extrinsic and 
intrinic sources of low-temperature magnetism. Our measurements were performed on a mosaic of hundreds of randomly oriented small aluminum (Al) flux-grown SmB$_6$ single crystals, and on a 
large doubly-isotope enriched $^{154}$Sm$^{11}$B$_6$ single crystal grown by the floating zone (FZ) method. 
The latter is the same $^{154}$Sm$^{11}$B$_6$ single crystal in which a 14~meV spin exciton has been detected by INS \cite{Alekseev:93,Alekseev:95,Fuhrman:15}. 

Flux-grown SmB$_6$ single crystals are known to contain Al inclusions \cite{Phelan:16}.   
Pure Al does not contain electronic moments, and while muons landing in the Al inclusions may sense static nuclear dipole
fields, these are decoupled on the same field scale as the B nuclear moments in our LF experiments. 
Samarium (Sm) vacancies, which act as ``Kondo holes'' in the strongly-correlated state of SmB$_6$, are more prevalent in FZ-grown single crystals \cite{Phelan:16}. 
Theoretically, a finite concentration of Sm vacancies introduces an impurity band in the hybridization gap and gives rise to a Curie-Weiss-like susceptibility \cite{Schlottmann:92}. 
They also adversely affect spin excitons, as evidenced by a Sm-vacancy induced suppression of the 16-18~meV exciton feature observed by Raman spectroscopy \cite{Valentine:16}.

\begin{figure}
\centering
\includegraphics[width=8.0cm]{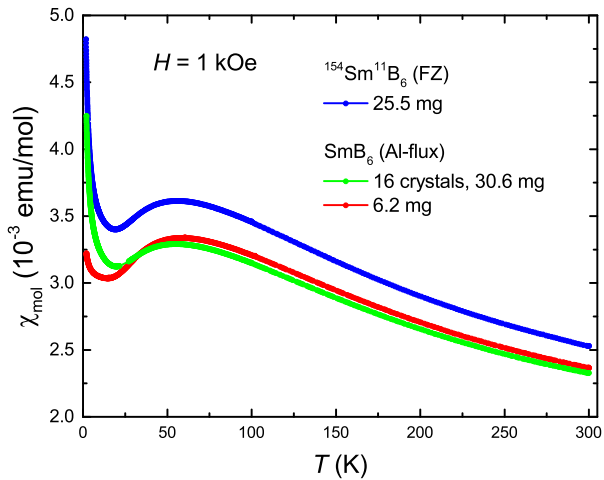}
\caption{(Color online) Temperature dependence of the bulk magnetic susceptibility at $H \! = \! 1$~kOe for a piece of 
the FZ-grown $^{154}$Sm$^{11}$B$_6$ single crystal, a mosaic of 16 of the Al-flux grown SmB$_6$ single crystals, and 
one of the Al-flux grown single crystals studied in Ref.~\cite{Akintola:18}.}
\label{fig1}
\end{figure}

Figure~\ref{fig1} shows representative $\chi(T)$ data for the two samples studied here and for one of the Al-flux grown single crystals
investigated in Ref.~\cite{Akintola:18}. No magnetic hysteresis was found in any of the samples. The magnetic susceptibility over much of the 
temperature range is a sum of contributions from the $4f^6$ Sm$^{2+}$ and $4f^55d^1$ Sm$^{3+}$ ion configurations \cite{Nickerson:71}.
There is a pronounced low-$T$ upturn in $\chi(T)$ for the current samples, and the overall susceptibility is greater in the larger $^{154}$Sm$^{11}$B$_6$ 
single crystal. Both features are clearly of extrinsic origin. 

\begin{figure}
\centering
\includegraphics[width=8.5cm]{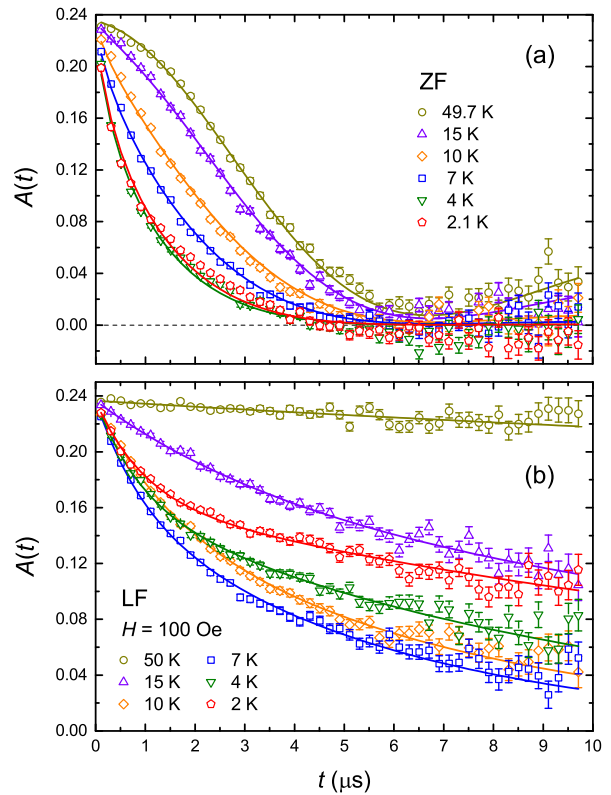}
\caption{(Color online) Representative (a) ZF- and (b) LF-$\mu$SR asymmetry spectra recorded on the $^{154}$Sm$^{11}$B$_6$ single crystal. The LF-$\mu$SR spectra
were recorded for a field $H_{\rm LF} \! = \! 100$~Oe applied parallel to the initial muon spin polarization. The solid curves through the data points 
are fits to Eq.~(\ref{eqn:Asy}).}
\label{fig2}
\end{figure}

Figure~\ref{fig2} shows typical ZF- and weak LF-$\mu$SR asymmetry spectra for the $^{154}$Sm$^{11}$B$_6$ single crystal, which are reasonably described by
\begin{equation}
A(t) = a_0 G_{\rm KT}(\Delta, t, H_{\rm LF}) e^{-[\lambda(T) t]^\beta} \, ,
\label{eqn:Asy}
\end{equation}
where $G_{\rm KT}(\Delta, t, H_{\rm LF})$ is the static Gaussian Kubo-Toyabe function \cite{Hayano:79} intended to account for the temperature-independent relaxation 
caused by the nuclear moments. It assumes a Gaussian field distribution of width $\Delta/\gamma_\mu$ (where $\gamma_\mu/2 \pi$ is the muon 
gyromagnetic ratio) and is dependent on the applied longitudinal field $H_{\rm LF}$. The LF-$\mu$SR spectra in Fig.~1(b) were recorded for $H_{\rm LF} \! = \! 100$~Oe, which is 
sufficient to completely decouple the muon spin from the nuclear dipole fields. The stretched-exponential function in Eq.~(\ref{eqn:Asy}) 
accounts for additional sources of magnetic field in the sample.  
Global fits of the ZF and 100~Oe LF spectra assuming $\beta$ is independent of temperature, yield $\beta \! = \! 0.562(5)$ and 0.552(9)
for the Al-flux grown sample, and $\beta \! = \! 0.699(2)$ and 0.658(3) for the FZ-grown single crystal. 
The ZF fits also yield $\Delta \! = \! 0.2336(6)$~$\mu$s$^{-1}$ and $\Delta \! = \! 0.2589(7)$~$\mu$s$^{-1}$ for the Al-flux and FZ grown samples, respectively.
The ZF values of $\Delta$ and $\beta$ are somewhat different from those obtained in previous $\mu$SR studies of SmB$_6$ \cite{Biswas:14,Biswas:17,Akintola:17},
highlighting variations in sample quality.

\begin{figure}
\centering
\includegraphics[width=8.5cm]{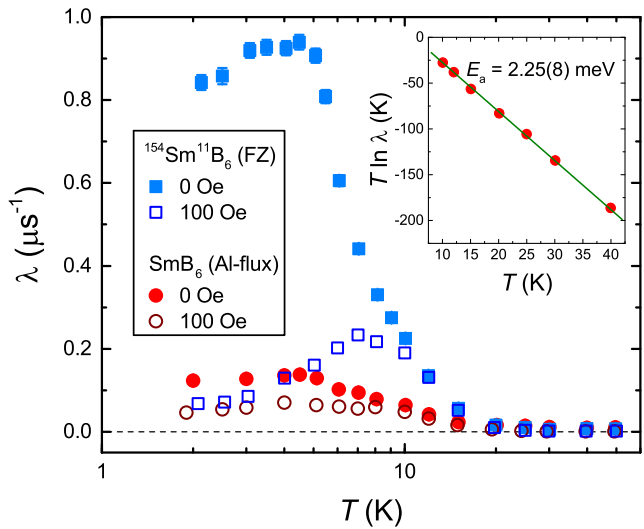}
\caption{(Color online) Temperature dependence of the ZF-$\mu$SR (solid symbols) and 100~Oe LF-$\mu$SR (open symbols) relaxation rate $\lambda$ for
the FZ-grown $^{154}$Sm$^{11}$B$_6$ and Al-flux grown SmB$_6$ single crystals. Inset: Temperature dependence of the ZF data for the Al-flux grown sample
at $T \! \ge \! 10$~K, shown as an Arrhenius plot in the form $T \ln \! \lambda$ vs $T$. The green line is a linear fit with intercept $E_{a}/k_{b}$.} 
\label{fig3}
\end{figure}
 
The temperature dependence of the fitted values of $\lambda$ for the two samples is shown in Fig.~\ref{fig3}. Below $\sim \! 20$~K, the ZF 
value of $\lambda$ increases more rapidly in the $^{154}$Sm$^{11}$B$_6$ single crystal. This behavior is qualitatively 
similar to previous findings \cite{Biswas:14}, although the difference between the FZ and Al-flux grown single crystals here is more extreme.
In the earlier ZF-$\mu$SR studies, a broad peak in $\lambda(T)$ was observed near 4 to 5~K \cite{Biswas:14,Biswas:17,Akintola:17}. The sharpness of this feature, 
however, is sample dependent. Here both samples display a maximum in $\lambda(T)$ for ZF near 4.5~K. Below $\sim \! 10$~K, the 100~Oe LF and ZF values of $\lambda(T)$ diverge
in both samples. The significant reduction of $\lambda$ by the 100~Oe field indicates the gradual development of weak local 
quasi-static magnetic fields as the temperature is lowered toward 2~K. In what follows, we demonstrate via LF-$\mu$SR results up to 4~kOe 
that this magnetism is dependent on the sample preparation. 

\begin{figure}
\centering
\includegraphics[width=8.5cm]{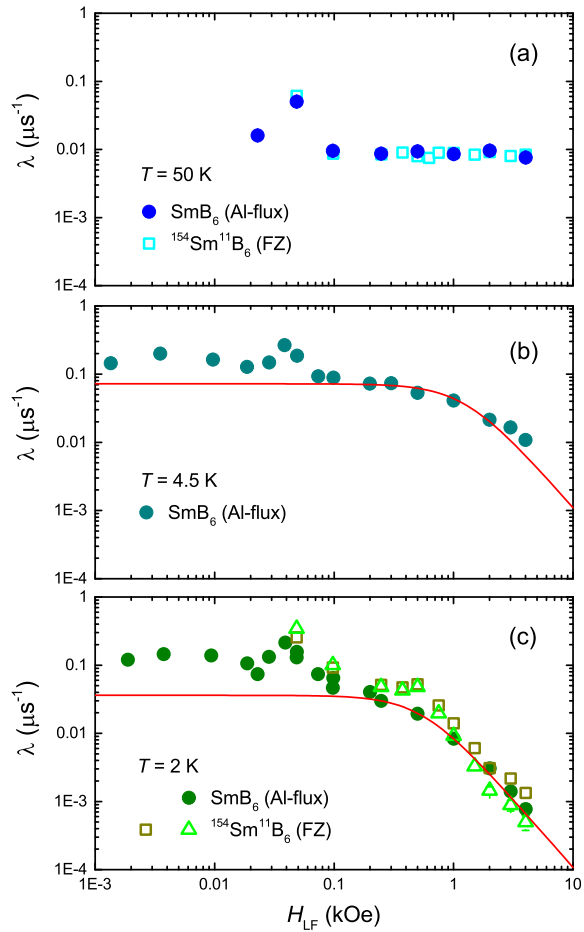}
\caption{(Color online) Field dependence of the relaxation rate $\lambda$ obtained from fits of the LF-$\mu$SR asymmetry spectra at (a) 50~K, (b) 4.5~K, and (b) 2~K.
Note that $\lambda$ is a ``pure'' exponential relaxation rate in (a), but a stretched-exponential relaxation rate in (b) and (c). The solid green circles and
open brown squares in (c) are from fits assuming the ZF values $\beta \! = \! 0.562$ and $\beta \! = \! 0.699$, respectively. The open green triangles are 
results for the $^{154}$Sm$^{11}$B$_6$ single crystal from fits assuming the ZF value $\beta \! = \! 0.562$ for the Al-flux grown sample.     
The solid red curves in (b) and (c) are fits of the $H_{\rm LF} \! > \! 100$~Oe data for the Al-flux grown single crystals to Eq.~(\ref{eqn:Redfield}).} 
\label{fig4}
\end{figure}

Above $\sim \! 20$~K we find the stretched-exponential relaxation function in Eq.~(\ref{eqn:Asy}) may be replaced by a pure exponential. 
Figure~\ref{fig4}(a) shows $\lambda(H_{\rm LF})$ at 50~K for both samples, obtained from an analysis with $\beta \! = \! 1$.
Biswas {\it et al.} \cite{Biswas:14} previously showed that the relaxation rate $\lambda(H_{\rm LF})$ below $H_{\rm LF} \! \sim \! 100$~Oe exhibits a 
broad peak centered near 40~Oe due
to an avoided level crossing resonance (ALCR) --- presumably due to a matching of the Zeeman splittings of the muon and B nuclear spins. 
While the influence of the ALCR is evident below 100~Oe, at higher field $\lambda(H_{\rm LF})$ is 
independent of field and identical in the two samples. Moreover, the average value of $\lambda(H_{\rm LF})$ between 100~Oe and 4~kOe is in 
good agreement with the ZF values of $\lambda$ at 50~K in Fig.~\ref{fig3}. Thus it is clear that the muons sense fast fluctuating internal fields of a similar rate 
in both samples at 50~K.  

Figures~\ref{fig4}(b) and (c) show $\lambda(H_{\rm LF})$ at 4.5~K and 2~K obtained from fits to Eq.~(\ref{eqn:Asy}) with $\Delta$ and $\beta$ fixed to the 
values determined from the analysis of the ZF-$\mu$SR spectra for each sample. In addition, we show results for the $^{154}$Sm$^{11}$B$_6$ single crystal at 2~K
from fits assuming the Al-flux grown value $\beta \! = \! 0.562$. Above 100~Oe there is good agreement between $\lambda(H_{\rm LF})$ for the two samples
and the data is well described by the Redfield formula \cite{Schenck:1985}       
\begin{equation}
\lambda(H_{\rm LF}) \! = \! \frac{\lambda(H_{\rm LF} = 0)}{1 + \left(\gamma_\mu H_{\rm LF} \tau \right)^2} \, ,
\label{eqn:Redfield}
\end{equation}
where $\lambda(H_{\rm LF} \! = \! 0) \! = \! 2 \gamma_\mu^2 \langle B_{\rm loc}^2 \rangle \tau$ and $\langle B_{\rm loc}^2 \rangle$ is the mean of the square of the 
transverse components of a local magnetic field fluctuating at a rate $1/\tau$. Equation~(\ref{eqn:Redfield}) is strictly valid for fast field fluctuations 
in a Gaussian distribution with a single fluctuation rate $1/\tau$, whereas a stretched-exponential relaxation often signifies a distribution of fluctuation 
rates. Nevertheless, the Redfield equation is adequate for achieving an approximate quantitative understanding of the data, provided the applied
field does not modify the magnetic fluctuation spectrum. 
A fit of the $\lambda(H_{\rm LF} \! > \! 100$~Oe) data 
for the Al-flux grown single crystals to Eq.~(\ref{eqn:Redfield}) yields $\lambda(H_{\rm LF} = 0) \! = \! 0.0361(6)$~$\mu$s$^{-1}$, $\tau \! = \! 2.15(6) \times 10^{-8}$~s
and $B_{\rm loc} \! = \! 10.8(5)$~G at 2~K, and $\lambda(H_{\rm LF} = 0) \! = \! 0.072(4)$~$\mu$s$^{-1}$, $\tau \! = \! 0.9(1) \times 10^{-8}$~s
and $B_{\rm loc} \! = \! 23(5)$~G at 4.5~K. We note that similar values are obtained from fits where $\beta$ is free to vary with $H_{\rm LF}$ \cite{Supplemental}.
The fitted value of $\lambda(H_{\rm LF} \! = \! 0)$ at 2~K is nearly 3.5 times smaller than the ZF value of $\lambda$
for the Al-flux grown SmB$_6$ single crystals, and $\sim \! 23$ times smaller than the ZF value of $\lambda$ for the FZ-grown $^{154}$Sm$^{11}$B$_6$ single crystal
(see Fig.~\ref{fig3}). This implies that a weak LF completely decouples the muon spin from a source of bulk magnetism, distinct from that of the nuclear
moments. Since the difference between the fitted value of $\lambda(H_{\rm LF} \! = \! 0)$ and the ZF value of $\lambda$ is much greater for the $^{154}$Sm$^{11}$B$_6$ single crystal,
the magnetism is likely due to a greater concentration of Sm vacancies and perhaps rare-earth impurities.
Contrarily, the similarity of $\lambda(H_{\rm LF})$ for the two samples above 100~Oe indicates that there is 
at least one other source of intrinsic bulk magnetism, which at 2~K gives rise to fluctuating magnetic fields of frequency on the order of $10^7$~Hz.

In the absence of a low-energy spin exciton, the existence of the intrinsic magnetism is surprising.
The implanted positive muon ($\mu^+$) senses the localized Sm-$4f$ moments via the magnetic dipole interaction and through an indirect RKKY 
interaction that spin polarizes the conduction electrons at the muon site. In zero field, the total field at the $\mu^+$ site ${\bf B}_{\mu}$ is the vector sum of the 
corresponding dipolar (${\bf B}_{\rm dip}$) and hyperfine contact (${\bf B}_{\rm hf}$) fields. The opening of the Kondo gap in SmB$_6$ is complete 
by $T \! \sim \! 30$~K \cite{Xu:14a,Zhang:13}. Consequently, ${\bf B}_{\rm dip}$ and ${\bf B}_{\rm hf}$ are expected to vanish at $T \! << \! 30$~K, due to complete screening of the 
Sm-$4f$ moments by the conduction electrons and the absence of a screening cloud of conductions electrons about the $\mu^+$.

If the intrinsic magnetism is associated with populating a low-energy spin exciton state, 
the ZF relaxation rate should obey an Arrhenius law $\lambda \! = \! \lambda_0 \exp(E_{\rm a}/k_{\rm B}T) \! \propto \! \tau$. 
Unfortunately, any such behavior for the $^{154}$Sm$^{11}$B$_6$ single crystal is masked by the large Sm vacancy/impurity contribution.
The ZF relaxation rate for the Al-flux grown sample does exhibit an Arrhenius behavior for $T \! \ge \! 10$~K, characterized by an activation energy
$E_{\rm a} \! = \! 2.25(8)$~meV (Fig.~\ref{fig3}, inset). Combined with the lower value $E_{\rm a} \! \sim \! 1$~meV determined from
60~kOe transverse-field $\mu$SR measurements of the relaxation rate in similar Al-flux grown single crystals \cite{Akintola:18}, 
these results are compatible with the field-dependent contribution to the $^{11}$B NMR spin-lattice relaxation rate --- which has been explained 
by in-gap magnetic states separated from the conduction band by a 2.6~meV gap that shrinks with increasing field and closes by 140~kOe \cite{Caldwell:07}. 
An $\sim \! 2.6$~meV zero-field gap has also been observed by magnetotransport measurements \cite{Shahrokhvand:17}, and in the low-energy electrodynamic 
response spectra of SmB$_6$ in the far-infrared range \cite{Gorshunov:99}. A 2.6~meV magnetic exciton is predicted to arise from a competition between 
the magnetic $4f^5$ and non-magnetic $4f^6$ multiplets \cite{Shick:15}. Based on solutions of the single-site Anderson impurity model, 
1.74~$\mu_B$ local magnetic moments are generated via excitation of a triplet state situated 2.6~meV in energy above an intermediate 
valence non-magnetic singlet ground state. However, one then expects a sharp neutron energy transfer peak at 2.6~meV or a dispersive peak 
associated with a collective mode of the localized moment system due to intersite interactions. The $\sim \! 2.6$~meV magnetic excitation 
may instead be connected to recent theoretical work that shows donor impurities in SmB$_6$ produce a mid-gap impurity band with a corresponding 
ionization energy of 1 to 5~meV \cite{Skinner:19}.

Below 4.5~K, the LF results clearly show that spin freezing in SmB$_6$ has an extrinsic origin
and there exist weak intrinsic fields ($B_{\rm loc} \! \sim \! 10.8$~G) that fluctuate too slow ($1/\tau \! \sim \! 47$~MHz) to originate 
from a magnetic excitation gap on the order of 1~meV.
The latter is similar to $\mu$SR findings in the Kondo insulator YbB$_{12}$, which indicate the presence of slowly fluctuating ($1/\tau \! \sim \! 60$~MHz)
weak internal fields ($B_{\rm loc} \! = \! 5.4$~G) below $\sim \! 5$~K \cite{Yaouanc:99}. The values of $B_{\rm loc}$
are consistent with very small localized Sm and Yb magnetic moments ($\sim \! 10^{-2} \mu_{\rm B}$). Since there is no further change in $\lambda$ down to millikelvin 
temperatures \cite{Biswas:14,Akintola:17,Yaouanc:99}, the small magnetic moment may indicate that the carrier density in these compounds
is insufficient to completely Kondo screen the localized $4f$ magnetic moments. On the other hand, $\mu$SR experiments on CaB$_6$ and BaB$_6$ show the
emergence of small random magnetic moments ($\sim \! 10^{-2} \mu_{\rm B}$/B) below $\sim \! 130$~K \cite{Kuroiwa:06}, perhaps associated 
with intrinsic defects detected in transport measurements of alkaline-earth-metal hexaborides \cite{Stankiewicz:16}. Here 
the similarity of the data for the FZ and Al-flux grown samples in Fig.~\ref{fig4}(c) seems to rule out Sm vacancies,
but the weakly dynamic small-moment magnetism may originate from other kinds of intrinsic defects.
    
In summary, we have shown there are multiple sources of low-$T$ magnetism in SmB$_6$.
In addition to underlying persistent slowly fluctuating weak moments, our experiments provide additional support for a bulk $\sim \! 2.6$~meV magnetic excitation.
The results on the neutron $^{154}$Sm$^{11}$B$_6$ sample demonstrate that enhanced Sm vacancies do not explain why a magnetic excitation of
this energy is not observed by INS. Since optical conductivity experiments link the $\sim \! 2.6$~meV excitation to charge carrier 
localization \cite{Gorshunov:99}, the origin might be connected to intrinsic defects and the creation of very small magnetic moments --- such that the neutron 
cross section is below the noise floor even in the large $^{154}$Sm$^{11}$B$_6$ sample.  
A very weak broad ``hump'' seen in the specific heat of a FZ-grown isotopically-enriched SmB$_6$ sample between 4~K to 10~K \cite{Orendac:17}
may be a manifestation of this magnetic excitation, as this feature is also observed in the specific heat of our samples. However, in contrast to the strong 
magnetic field dependence of the $^{11}$B NMR $1/T_1$ maximum and $\mu$SR relaxation rate in this temperature range, the field dependence of 
this very small specific heat hump is almost negligible up to 140~kOe. Lastly, we note that there is a similar 2.7~meV electronic excitation \cite{Gorshunov:06} 
and an analogous field-dependent $^{11}$B NMR $1/T_1$ maximum \cite{Shishiuchi:02} in YbB$_{12}$, suggesting the sources of the intrinsic low-$T$ magnetism in 
these two candidate TKIs are the same. 

\begin{acknowledgments}
We thank P.S. Riseborough and A.B. Shick for informative discussions. 
J.P. acknowledges support from the Gordon and Betty Moore Foundation's EPiQS Initiative through Grant No. GBMF4419, and the National Institute of Standards and 
Technology Cooperative Agreement 70NANB17H301. W.T.F. is grateful to the ARCS foundation and the Schmidt Science Fellows program, 
in partnership with the Rhodes Trust, for the partial support of this work. J.E.S. acknowledges support from the Natural Sciences and 
Engineering Research Council of Canada. We thank P.A. Alekseev and J.-M. Mignot for providing access to the $^{154}$Sm$^{11}$B$_6$ single crystal.
\end{acknowledgments}


\end{document}